\documentclass[11pt,twoside]{article}
\usepackage[pdftex]{graphicx}
\usepackage{amsmath}
\usepackage{amssymb}
\usepackage{cite}

\DeclareMathOperator{\ctg}{ctg}
\DeclareMathOperator{\const}{const}
\newcommand{\w}{\mathit w}

\begin{document}

\newcommand{\pst}{\hspace*{1.5em}}

\newcommand{\be}{\begin{equation}}
\newcommand{\ee}{\end{equation}}
\newcommand{\bm}{\boldmath}
\newcommand{\ds}{\displaystyle}
\newcommand{\bea}{\begin{eqnarray}}
\newcommand{\eea}{\end{eqnarray}}
\newcommand{\ba}{\begin{array}}
\newcommand{\ea}{\end{array}}
\newcommand{\arcsinh}{\mathop{\rm arcsinh}\nolimits}
\newcommand{\arctanh}{\mathop{\rm arctanh}\nolimits}
\newcommand{\bc}{\begin{center}}
\newcommand{\ec}{\end{center}}

\thispagestyle{plain}

\label{sh}


\begin{center} {\Large \bf
\begin{tabular}{c}
THE DRIVEN-OSCILLATOR EVOLUTION
\\[-1mm]
IN THE TOMOGRAFIC-PROBABILITY REPRESENTATION
\end{tabular}
 } \end{center}

\bigskip

\bigskip

\begin{center} {\bf
Dmitry B. Lemeshevskiy$^{1*}$ and Vladimir I. Man'ko$^{2}$  
}\end{center}

\medskip

\begin{center}
{\it
$^1$Moscow Institute of Physics and Technology (State University)\\
Institutskii per. 9, Dolgoprudnyi, Moscow Region 141700, Russia
\smallskip

$^2$P. N. Lebedev Physical Institute, Russian Academy of Sciences\\
Leninskii Prospect 53, Moscow 119991, Russia
}
\smallskip

$^*$Corresponding author e-mail:~~~d.lemeshevskiy@gmail.com,~~~manko@sci.lebedev.ru\\
\end{center}

\begin{abstract}\noindent
We consider the problem of the driven harmonic oscillator in the probability representation of quantum
mechanics, where the oscillator states are described by fair nonnegative probability distributions of position
measured in rotated and squeezed reference frames in the system's phase space. For some specific
oscillator states like coherent states and nth excited states, the tomographic-probability distributions
(called the state tomograms) are found in an explicit form. The evolution equation for the tomograms
is discussed for the classical and quantum driven oscillators, and the tomographic propagator for this
equation is studied.
\end{abstract}

\medskip


\section{Introduction}
\pst Recently \cite{ManPLA96, IbortPhysScr2009, ManManFoundPhys2011, ManManJRLR97} the new formulation of quantum mechanics based on identification of any quantum
state with the tomographic-probability distribution containing the same information on the state that
is given by the state wave function \cite{Schr} or the density operator \cite{Landau, vonNeum}was introduced. This formulation
is the result of many previous attempts to find such a notion of the quantum state which is similar to
the notion of classical state associated with the probability density of the classical system in its phase
space $(q,p)$. These attempts provided the notion of the Wigner function $W(q, p)$ introduced in \cite{Wign32}, the Husimi function $Q(q,p)$ introduced in \cite{Hus40}, as well as Glauber function $P(q,p)$ \cite{Glau63} and Sudarshan function $\varphi (z)$ \cite{Sudar63}. All these functions were associated with system density operator by invertable maps. Also all these functions are similar in some aspects to classical probability distribution $f(q,p)$. Nevertheless all these functions, called quasiprobability distributions or quasiprobabilities, are not fair probability distributions in the phase space.\\
\pst At the same time, optical tomogramm $\w (X,\theta)$, introduced as a technical tool \cite{BerBer, VogRis} and related to the Wigner function by the integral Radon transform \cite{Rad} is a measurable in quantum optics experiments \cite{Raym93} probability distribution of the photon homodyne quadrature $X$, where the angle $\theta$ is the local oscillator phase. Optical tomogramms were used to reconstruct the Wigner function (considered as a concept of quantum state) from the experimental data. In \cite{ManPLA96} the notion of the symplectic tomogramm $\w (X,\mu,\nu)$ was suggested. It is also a probability distribution of the photon quadrature $X$, related with optical tomogramm by an invertible map. Also it became clear that tomographic probabilities can be considered as primary objects (not as technical tools) and can be identified with quantum states. These tomogramms are alternative to the wave function and density operator since they contain complete information about the quantum states.\\
\pst The problem of driven oscillator is an interesting problem studied first in \cite{Hus53, Schw60, Feynm48}. The coherent states of the oscillator and dynamical invariants \cite{Suslov} of such system were found in \cite{TrifPRD70, MalMan, DodMan183}. On the other hand, a detail consideration of this problem in the tomographic probability representation has never been performed. The aim of our work is to study driven harmonic oscillator in the tomographic probability representation. We obtained the symplectic and optical tomogramms as solutions of the tomographic evolution equation for coherent and excited states of driven oscillator.\\
\pst This paper is organized as follows.\\
\pst We find the Green function for the Schr\"odinger equation in Sec. 2 and the evolution equations for
the classical probability distribution in Sec. 3, where we also obtain the Wigner function and classical
and quantum tomograms. We study symplectic and optical tomograms of coherent and excited states in
Secs. 4 and 5, and finally we give our conclusions in Sec. 6.

\section{Green Function of Driven Harmonic Oscillator}
\pst
We study the problem of a quantum oscillator with a time-dependent external force. The Hamiltonian
of such a system reads (we put $m =1, \omega = 1, \hbar = 1$):
\begin{equation}
\label{hamiltonian}\hat H = \frac{\hat q^2}2+\frac{\hat p^2}2-f(t)\hat q
\end{equation}
Our goal is to find the expression for Green's function $ G(x, x', t)$ of Schr\"odinger evolution equation.For this, we
employ the method of integrals of motion \cite{Suslov, TrifPRD70, MalMan, DodMan183}.We consider a classical analog of the Hamiltonian~(\ref{hamiltonian}):
\begin{equation}
\label{hamclas}H(q,p,t) = \frac{q^2}2+\frac{p^2}2-f(t)q
\end{equation}
The classical equations of motion determining the oscillator position and momentum read
\begin{gather}
\label{cl1}\dot q=p\\
\label{cl2}\dot p=-q+f(t)
\end{gather}
and the classical trajectories in the oscillator phase space under the initial conditions $q(0)=q_0, p(0)=p_0$ are given by formulae:
\begin{gather}
\label{q(t)}q(t)=q_0\cos t + p_0\sin t + \int^t_0\sin (t-\tau)f(\tau)d\tau\\
\label{p(t)}p(t)=p_0\cos t - q_0\sin t + \int^t_0\cos (t-\tau)f(\tau)d\tau
\end{gather}
Now we consider the system of equations~(\ref{q(t)})~and~(\ref{p(t)})~as an algebraic system for unknown initial position $q_0$ and momentum $p_0$, respectively. The variables $q,p,t$ are taken as parameters. The solution of the system is given as:
\begin{gather}
\label{q0}
q_0(q,p,t)= q\cos t-p\sin t +\int^t_0 \sin \tau f(\tau)d\tau=q\cos t-p\sin t + F(t)\\
\label{p0}
p_0(q,p,t)=  p\cos t+q\sin t -\int^t_0 \cos \tau f(\tau)d\tau =p\cos t+q\sin t - J(t)
\end{gather}
We define operators acting in the Hilbert space of the quantum driven oscillator states as follows:
\begin{gather}
\label{q0_d}\hat q_0\left(\hat q, \hat p, t\right)=\hat q\cos t-\hat p\sin t+F(t)\\
\label{p0_d}\hat p_0\left(\hat q, \hat p, t\right)=\hat q\sin t+\hat p\cos t-J(t)
\end{gather}
Calculating the total derivative of the operator $\hat q_0(q,p,t)$ with respect to time $t$ we obtain:
\begin{equation}
\frac{d\hat q_0}{dt} = \frac{\partial \hat q_0}{\partial t}+i\left[ \hat H, \hat q_0\right] =\dot F(t) - f(t)\sin t = 0
\end{equation}
where we take into account the definition of function $F(t)=\int^t_0 \sin \tau f(\tau)d\tau$ given by the~(\ref{q0}).

Similarly, the total time derivative of the operator $\hat p_0(q,p,t)$:
\begin{equation}
\frac{d\hat p_0}{dt} = \frac{\partial \hat p_0}{\partial t}+i\left[ \hat H, \hat p_0\right] =-\dot J(t) +f(t)\cos t = 0
\end{equation}
where the definition of function $J(t)=\int^t_0 \cos \tau f(\tau)d\tau$ from~(\ref{p0}) is taken into account. 

Thus, operators~(\ref{q0_d})~and~(\ref{p0_d}) are integrals of motion and correspond to initial position and momentum. Then these operators must satisfy the equations for the Green's function $ G(x, x', t)$ \cite{TrifPRD70, MalMan, DodMan183}: 
\begin{gather}
\label{q0G}\hat q_{0(x)}G(x, x', t)=\hat q_{(x')}G(x,x',t)\\
\label{p0G}\hat p_{0(x)}G(x, x', t)=-\hat p_{(x')}G(x,x',t)
\end{gather}
where the operators in the left-hand sides of equations act on variable $x$, and in the right-hand sides on $x'$. Now we write Eqs.~(\ref{q0G})~and~(\ref{p0G})~explicitly:
\begin{gather}
\label{1G}\left(x\cos t+i\frac{\partial}{\partial x}\sin t+F(t)\right)G(x,x',t)=x'G(x,x',t)\\
\label{2G}\left(x\sin t-i\frac{\partial}{\partial x}\cos t-J(t)\right)G(x,x',t)=i\frac{\partial}{\partial x'}G(x,x',t)
\end{gather}
where we still need to define function $\tilde x(t)$ to obtain more convenient equations,
\begin{equation}
\label{-x(t)}F(t)\cos t-J(t)\sin t= -\int^t_0 \sin (t-\tau)f(\tau)d\tau=-\tilde x(t)
\end{equation}
It is worth noting that $\tilde x(t)$ is solution of the classical equations of motion (\ref{cl1}), (\ref{cl2}) with zero initial conditions: $q(0)=0, p(0)=0$. \\
\pst
Finally, system of equations for defining Green's function $G(x,x',t)$ is: 
\begin{gather}
\label{G1f}\frac{\partial G}{\partial x}=\left(-\frac{x\cos t -x'}{i\sin t}-\frac{F(t)}{i\sin t}\right)G\\
\label{G2f}\frac{\partial G}{\partial x'}=\left(-\frac{x'\cos t -x}{i\sin t}-\frac{\tilde x(t)}{i\sin t}\right)G
\end{gather}
where we skipped the arguments of the Green function. 

Now one can integrate~(\ref{G1f})~with respect to variable $x$ and~(\ref{G2f})~with respect to variable $x'$. Afterperforming the integration, we arrive at
\begin{equation}
\label{Gxx}G(x,x',t)=C(t)\exp \left(-\frac{\left( x^2+x'^2\right)\cos t-2xx'}{2i\sin t}-\frac{xF(t)+x'\tilde x(t)}{i\sin t}\right)
\end{equation}
Finally, following the procedure adopted from \cite{Suslov, TrifPRD70, MalMan, DodMan183}, we obtain the equation for the function $C(t)$ as follows:
\begin{equation}
\label{Gschr}i\frac{\partial G\left( x,x',t \right)}{\partial t}=\hat H_{(x)}G(x,x',t)
\end{equation}
After some algebra, we obtain an equation that does not contain the variables $x,x'$ is obtained:
\begin{equation}
\label{dC(t)}i\frac{dC(t)}{dt}=C(t)\left(\frac{\ctg t}{2i}+\frac{F^2(t)}{2\sin^2 t}\right)
\end{equation}
Equation~(\ref{dC(t)})~can be simply integrated with respect to time $t$ and one obtains the function $C(t)$ in a form containig a constant $C$:
\begin{equation}
\label{C(t)}C(t)=\frac C{\sqrt {\sin t}}\exp\left(-\frac i2\int^t_0\frac{F^2(\tau)}{\sin^2 \tau}d\tau\right) 
\end{equation}
To calculate the constant $C$ we require that the following condition at the initial moment of time $t=0$ be satisfied:
\begin{equation}
\label{Gt0}G(x,x',t=0)=\delta (x-x')
\end{equation}
After calculating $C$ using~(\ref{Gt0})~and substituting it into~(\ref{C(t)})~and~(\ref{Gxx})~we arrive at the final expression for
the Green function of the Schr\"odinger evolution equation for the wave function,
\begin{multline}
\label{Gxxt}G(x,x',t)=\frac 1{\sqrt {2\pi i\sin t}}\exp\left(-\frac i2\int^t_0\frac{F^2(\tau)}{\sin^2 \tau}d\tau\right)\times\\ 
\times \exp \left(-\frac{\left( x^2+x'^2\right)\cos t-2xx'}{2i\sin t}-\frac{xF(t)+x'\tilde x(t)}{i\sin t}\right)
\end{multline}
\pst We consider some particular cases as examples.

We start with the case of a constant external force,$f(t)=f_0=\const$. In this particular case we calculate functions $\tilde x(t)$ and $F(t)$ according to their definitions~(\ref{-x(t)}) and~(\ref{q0}) correspondingly and substitute the result into the final expression~(\ref{Gxxt}).After some algebra, we obtain
\begin{equation}
G(x,x',t)=\frac 1{\sqrt {2\pi i \sin t}}\exp\left(\frac {if_0^2t}2-\frac{\left(\left(x-f_0\right)^2+\left(x'-f_0\right)^2\right)\cos t-2\left(x-f_0\right)\left(x'-f_0\right)}{2i\sin t}\right)
\end{equation}
Taking into account that Hamiltonian in this particular case differs from Hamiltonian of a harmonic oscillator with shifted equilibrium just in a constant term: $$\hat H = \dfrac 12\left(\hat p^2+\left(\hat q-f_0\right)^2\right)-\dfrac{f_0^2}2$$ we write the following set of equalities:
$$\hat H_{(x)}G(x,x',t)=\exp\left(\frac{if_0^2t}2\right)i\frac{\partial}{\partial t}\left(G(x,x',t)\exp\left(-\frac{if_0^2t}2\right)\right)-\frac{f_0^2}2G(x,x',t)=i\frac{\partial}{\partial t}G(x,x',t)$$
This means that Green function satisfies the Schr\"odinger evolution equation.\\
\pst
Now we considerthe limit case $\omega \to 0$. 

First of all we should recover $m, \omega, \hbar$ in formulae according to the following rules:
$$t\to\omega t; x\to\sqrt\frac{m\omega}{\hbar}x; f\to\frac f{\omega\sqrt{m\hbar\omega}}$$
After all the substitutions we have the Green function in the form
\begin{multline}
\label{Grin_w}G(x,x',t)=\\
=\sqrt{\frac{m}{2\pi i\hbar}}\sqrt{\frac{\omega}{\sin {\omega t}}}\exp\left[\frac{i}{2\omega^3m\hbar}\left(f_0^2\omega t+2f_0^2\frac{\cos{\omega t}-1}{\sin{\omega t}}-2f_0m\omega^2\left(x+x'\right)\frac{\cos{\omega t}-1}{\sin{\omega t}}+\right.\right.\\
+\left.\left.m^2\omega^4\frac{\left(x^2+x'^2\right)\cos{\omega t}-2xx'}{\sin{\omega t}}\right)\right]
\end{multline}
At $\omega\to 0$ the Green function reads:
\begin{equation}
G(x,x',t)=\sqrt{\frac{m}{2\pi i\hbar t}}\exp\left[\frac{i}{\hbar}\left(-\frac{f_0^2}{m}\frac{t^3}{24}+\frac{f_0t\left(x+x'\right)}{2}+\frac{m\left(x-x'\right)^2}{2t}\right)\right]
\end{equation}
By direct substitution into the Schr\"odinger evolution equationwe can check the correctness of our
calculations. It is also worth noting that the Green-function phase is the classical action $S(x,x',t)$, which satisfies the Hamilton-Jacobi equation, multiplied by  $\dfrac{i}{\hbar}$.

\section{Time Evolution of Classical and Quantum Distribution Functions}
\pst
Now we consider the propagators of different distribution functions of the system with Hamiltonian~(\ref{hamiltonian}) in the quantum case and Hamiltonian~(\ref{hamclas})in the classical case. We start from the classical distribution
function and the Liouville equation,
\begin{equation}
\label{rhoevol}\left(\frac{\partial}{\partial t}+p\frac{\partial}{\partial q}+(f(t)-q)\frac{\partial}{\partial p}\right)\rho (q,p,t)=0
\end{equation}
Noting that the system trajectories in the phase space are given by Eqs.~(\ref{q(t)}),~(\ref{p(t)}) (and, correspondingly, integrals of motion are $q_0(q, p, t), p_0(q, p, t)$ from~(\ref{q0}),~(\ref{p0})), we conclude that the evolution of the function $\rho (q,p,t)$ reads:
\begin{equation}
\label{rho(t)}\rho (q,p,t)=\rho_0(q\cos t-p\sin t+F(t), q\sin t+p \cos t -J(t))
\end{equation}
where $\rho_0(q,p)$ is the distribution function at initial moment of time $t=0$ and it depends on two variables $q, p$ only.\\
\pst
Now we consider the Moyal's equation \cite{Moyal49} for the Wigner's function $W(q,p,t)$:
\begin{equation}
\label{moyal}\left(\frac{\partial}{\partial t}+p\frac{\partial}{\partial q}-\frac{1}{i}\left[U\left(q\to q+\frac{i}{2}\frac{\partial}{\partial p}\right)-c.c.\right]\right)W(q,p,t)=0
\end{equation}
where the quantities in the square brackets are pure imaginary and in the equations, pure real ones. The
term in the square brackets can be rewritten, in view of the fact that the potential energy is given by $U(x)=\dfrac{x^2}{2}-f(t)x$, as follows:
\begin{equation}
U\left(q+\frac{i}{2}\frac{\partial}{\partial p}\right)=\Re U+\frac{i}{2}\frac{\partial}{\partial p}\left(q-f(t)\right)
\end{equation}
We are interested just in imaginary part of this expression. After substituting this into~(\ref{moyal})~we obtain the equation:
\begin{equation}
\label{dW(q,p,t)}\left(\frac{\partial}{\partial t}+p\frac{\partial}{\partial q}+(f(t)-q)\frac{\partial}{\partial p}\right)W(q,p,t)=0
\end{equation}
It is obvious that Eq.~(\ref{dW(q,p,t)}), being the evolution equation of the Wigner function $W(q,p,t)$ fully coincides with the equation~(\ref{rhoevol})~of evolution of the classical distribution function.The dependence of the Wigner
function on time is given by an expression similar to~(\ref{rho(t)}),
\begin{equation}
W(q,p,t)=W_0(q\cos t-p\sin t+F(t), q\sin t+p \cos t -J(t))
\end{equation}
\pst
Now we move to the tomographic-probability-distribution functions.

We start from classical tomogramm distribution. In thisl case we have \cite{MenMan99, ManManJRLR97} the following evolution equation:
\begin{equation}
\left(\frac{\partial}{\partial t}-\mu\frac{\partial}{\partial\nu}+\nu F\left(q\to-\left(\frac{\partial}{\partial X}\right)^{-1}\frac{\partial}{\partial\mu},t\right)\frac{\partial}{\partial X}\right)\w (X,\mu, \nu, t)=0
\end{equation}
where $F(q,t)$ has the meaning of acting force and in our case it is given by the expression $F(q,t)=-q+f(t)$. So, evolution equation in this case has following form:
\begin{equation}
\label{tomevol}\left(\frac{\partial}{\partial t}-\mu\frac{\partial}{\partial\nu}+\nu\frac{\partial}{\partial\mu}+\nu f(t)\frac{\partial}{\partial X}\right)\w (X,\mu, \nu, t)=0
\end{equation}
The integrals of the equation~(\ref{tomevol})~can be easily obtained. They read:
\begin{gather}
\label{mu0}\nu_0(X, \mu, \nu, t)=\nu\cos t+\mu\sin t\\
\mu_0(X, \mu, \nu, t)=\mu\cos t-\nu\sin t\\
\label{X0}X_0(X, \mu, \nu, t)=X-\mu\tilde x(t)-\nu\tilde p(t)
\end{gather}
where functions $\tilde x(t), \tilde p(t)$ are solutions of classical equations of motion~(\ref{cl1}),~(\ref{cl2}) with zero initial conditions. One should note that at the initial moment of time $t=0$:
$$\nu_0=\nu, \mu_0=\mu,X_0=X$$
hence, evolution of the tomographic-probability distribution is given by:
\begin{equation}
\label{w(t)}\w(X,\mu,\nu,t)=\w_0\left(X-\mu\tilde x(t)-\nu\tilde p(t), \mu\cos t-\nu\sin t, \nu\cos t+\mu\sin t\right)
\end{equation}
where $\w_0(X, \mu,\nu)$ is the tomographic-probability distribution at the initial moment of time and it depends just on three variables $X, \mu, \nu$.\\
\pst
We move now to the quantum tomogram and write the evolution equation according to\cite{ManPLA96, ManFoundPh97}:
\begin{equation}
\left(\frac{\partial}{\partial t}-\mu\frac{\partial}{\partial\nu}-i\left[U\left(q\to-\left(\frac{\partial}{\partial X}\right)^{-1}\frac{\partial}{\partial\mu}-i\frac{\nu}{2}\frac{\partial}{\partial X},t\right)-c.c.\right]\right)\w(X,\mu,\nu,t)=0
\end{equation}
We are interested just in the imaginary part of the expression in the square brackets. It reads $\nu\dfrac{\partial}{\partial\mu}+\nu f(t)\dfrac{\partial}{\partial X}$, and, finally, the evolution equation is
\begin{equation}
\left(\frac{\partial}{\partial t}-\mu\frac{\partial}{\partial\nu}+\nu\frac{\partial}{\partial\mu}+\nu f(t)\frac{\partial}{\partial X}\right)\w (X,\mu, \nu, t)=0
\end{equation}
that fully coincides with the equation in the classical case~(\ref{tomevol}). Therefore, evolution of quantum tomogramm is given by the same equation~(\ref{w(t)}).

\section{Symplectic Tomograms of Coherent and Excited States}
\pst
Now we study two examples of evolution of quantum tomogramm distributions. 

The first example is the coherent-state evolution, and the second example is the evolution of the nth
excited state of the harmonic oscillator.\\
\pst
The wave function of the harmonic-oscillator coherent state \cite{KlauSud, Glau63, Sudar63} at the initial moment of time can be written as follows:
\begin{equation}
\psi (x)=\frac{1}{\sqrt[4]{\pi}}\exp\left(-\frac{(x-x_0)^2}{2}+ip_0x\right)
\end{equation}
We calculate the tomographic-probability distribution of the harmonic-oscillator coherent state at the
initial moment of time using the following formula \cite{MenMan99}:
\begin{equation}
\label{psitow}\w (X,\mu,\nu)=\frac{1}{2\pi |\nu|}\left| \int\psi (y)\exp\left(\frac{i\mu}{2\nu}y^2-\frac{iX}{\nu}y\right)dy\right|^2
\end{equation}
After performing the integration, we arrive at
\begin{equation}
\label{w0coh}\w_0 (X,\mu,\nu)=\frac{1}{\sqrt\pi\sqrt{\mu^2+\nu^2}}\exp\left(-\frac{\left(X-\nu p_0-\mu x_0\right)^2}{\nu^2+\mu^2}\right)
\end{equation}
Now the time dependence of the tomographic-probability distribution is given, in view of~(\ref{w(t)}), as follows::
\begin{equation}
\w (X, \mu, \nu, t)=\frac{1}{\sqrt\pi\sqrt{\mu^2+\nu^2}}\exp\left(-\frac{\left(X-\mu x_{cl}(t)-\nu p_{cl}(t)\right)^2}{\nu^2+\mu^2}\right)
\end{equation}
where the functions $x_{cl}(t), p_{cl}(t)$ are solutions of classical equations of motion~(\ref{cl1}),~(\ref{cl2}) under the initial conditions $x(0)=x_0, p(0)=p_0$. \\
\pst
Now we consider the evolution of the nth excited state of the harmonic oscillator with the wave
function
\begin{equation}
\psi_n(x)=\frac{1}{\sqrt{2^nn!\sqrt{\pi}}}e^{-x^2/2}H_n(x)
\end{equation}
where $H_n(x)=(-1)^ne^{x^2}\dfrac{d^n}{dx^n}e^{-x^2}$ is Hermite polynomial. We show two ways of finding the evolution
of the system --- using the propagator for the wave function (Green function $G(x,x',t)$ from~(\ref{Gxxt})) and the evolution of the quantum tomogram given by~(\ref{w(t)}).\\
\pst
To study the evolution of the wave function, we start from the integral
\begin{equation}
\label{In}I_n=\int\exp\left[\left(-\frac{1}{2}+\alpha\right)x^2+\beta x\right]H_n(x)dx
\end{equation}
where $\alpha$ is a pure imaginary quantity. As a result, we obtained \cite{GradRyz}:
\begin{equation}
\label{In}I_n=\sqrt{\frac{2\pi}{1-2\alpha}}\exp\left(\frac{\beta^2}{2(1-2\alpha)}\right)i^n\left(\frac{1+2\alpha}{\sqrt{1-4\alpha^2}}\right)^nH_n\left(-\frac{i\beta}{\sqrt{1-4\alpha^2}}\right)
\end{equation}
and for the evolution of the wave function
\begin{multline}
\label{psi(x,t)}\psi(x,t)=\int G(x,x't)\psi(x',t=0)dx'=\frac{e^{-i(n+1/2)t}}{\sqrt{2^nn!\sqrt{\pi}}}\times\\
\times\exp\left(-\frac{i}{2}\left(\int^t_0\frac{F^2(\tau)}{\sin^2\tau}d\tau+\tilde x^2(t)\ctg t\right)\right)\exp\left(-\frac{(x-\tilde x(t))^2}{2}+i\tilde p(t)x\right)H_n(x-\tilde x(t))
\end{multline}
Now we use formula~(\ref{psitow}) to obtain the expression for the tomographic-probability distribution at  any particular moment of time $t$ from expression for the wave function~(\ref{psi(x,t)}):
\begin{equation}
\label{wn(t)}\w_n(X,\mu,\nu,t)=\frac{1}{2^nn!\sqrt{\pi}}\frac{1}{\sqrt{\nu^2+\mu^2}}\exp\left(-\frac{(X-\mu\tilde x(t)-\nu\tilde p(t))^2}{\nu^2+\mu^2}\right)H_n^2\left(\frac{X-\mu\tilde x(t)-\nu\tilde p(t)}{\sqrt{\nu^2+\mu^2}}\right)
\end{equation}
\pst
Finally, we show the second way of calculations.

The tomographic-probability distribution at the initial moment of time $t=0$ according to formula~(\ref{psitow}) reads:
\begin{equation}
\label{w0n}\w_n(X,\mu,\nu,t=0)=\frac{1}{2^nn!\sqrt{\pi}}\frac{1}{\sqrt{\nu^2+\mu^2}}\exp\left(-\frac{X^2}{\nu^2+\mu^2}\right)H_n^2\left(\frac{X}{\sqrt{\nu^2+\mu^2}}\right)
\end{equation}
For the nth excited state of the harmonic oscillator, the mean values of the position and momentum are
equal to zero, so the initial conditions for the functions $x_{cl}(t), p_{cl}(t)$ are zero and these functions coincide with $\tilde x(t), \tilde p(t)$. Then using formula for the evolution of the tomographic-probability distribution~(\ref{w(t)})~we obtain the same time dependence of the tomogramm distribution~(\ref{wn(t)}).

\section{Optical Tomograms of the Driven Oscillator}
\pst
In this section, we present all the results obtained in the previous sections in terms of the optical
tomogram $\w (X,\theta)$.The evolution equation for the optical tomogram can be derived from the Moyal
equation~(\ref{moyal})~using folowing correspondence rules\cite{KorJRLR2011}:
\begin{gather}
qW(q,p)\longleftrightarrow\left(\left(\frac{\partial}{\partial X}\right)^{-1}\sin\theta\frac{\partial}{\partial\theta}+X\cos\theta\right)\w (X,\theta)\\
pW(q,p)\longleftrightarrow\left(-\left(\frac{\partial}{\partial X}\right)^{-1}\cos\theta\frac{\partial}{\partial\theta}+X\sin\theta\right)\w (X,\theta)\\
\frac{\partial}{\partial q}W(q,p)\longleftrightarrow\cos\theta\frac{\partial}{\partial X}\w (X,\theta)\\
\frac{\partial}{\partial p}W(q,p)\longleftrightarrow\sin\theta\frac{\partial}{\partial X}\w (X,\theta)
\end{gather}
After such substitutionswe can write Eq.~(\ref{dW(q,p,t)})~in the form containing the optical tomogram $\w (X, \theta, t)$:
\begin{equation}
\label{wopt}\left(\frac{\partial}{\partial t}-\frac{\partial}{\partial\theta}+f(t)\frac{\partial}{\partial X}\sin\theta\right)\w (X, \theta,t)=0
\end{equation}
Integrals of motion for Eq.~(\ref{wopt})~can be written in a form, similar to formulae~(\ref{mu0})-(\ref{X0}):
\begin{gather}
\label{theta0}\theta_0(X, \theta, t)=\theta+t\\
\label{X0opt}X_0(X, \theta, t)=X-\tilde x(t)\cos\theta-\tilde p(t)\sin\theta
\end{gather}
In view of~(\ref{theta0})~and~(\ref{X0opt}) we obtain the final expression for the time evolution of the quantum optical tomogram
of the driven harmonic oscillator
\begin{equation}
\w (X,\theta,t)=\w_0(X-\tilde x(t)\cos\theta-\tilde p(t)\sin\theta, \theta+t)
\end{equation}
where $\w_0(X,\theta)$ is optical tomogramm distribution at the initial moment of time $t=0$ and it depends on just two variables $X, \theta$.\\
\pst
Optical tomograms at the initial moment of time $\w_0(X,\theta)$ for the cases considered in previous section can be found from~(\ref{w0coh}),~(\ref{w0n}) using the substitution $ \mu\longrightarrow\cos\theta,\nu\longrightarrow\sin\theta$.
So, for the
coherent state of the driven oscillator the time evolution of the optical tomogram reads
\begin{equation}
\w (X,\theta,t)=\frac{1}{\sqrt\pi}\exp\left( -\left(X-x_{cl}(t)\cos\theta-p_{cl}(t)\sin\theta\right)^2\right)
\end{equation}
In the case of the nth excited state of the harmonic oscillator, the evolution of the optical tomogram is
\begin{equation}
\w_n (X,\theta,t)=\frac{1}{2^nn!\sqrt\pi}\exp\left(-\left(X-\tilde x(t)\cos\theta-\tilde p(t)\sin\theta\right)^2\right)H_n^2\left(X-\tilde x(t)\cos\theta-\tilde p(t)\sin\theta\right)
\end{equation}

\section{Conclusions}
\pst
To conclude, we point out the main results of our study.

We presented a short review of the tomographic probability representation of quantum mechanics on
the example of the driven harmonic oscillator. We considered the evolution equation for the tomographicprobability
densities of the symplectic and optical tomograms. We concentrated on the important solutions
of the evolution equations which correspond to the Gaussian packets (normal distributions of the
coherent state) and Gauss–Hermite distributions corresponding to the oscillator excited states. Other
tomographic approaches to the driven-oscillator problem like the photon-number tomographic scheme
will be studied in future publications.

\end{document}